\documentclass[aps,pra,showpacs,twocolumn]{revtex4}
\usepackage{graphicx}
\usepackage{bm}
\usepackage{amssymb}

\begin{document}
\title{Upper bounds on success probabilities in linear optics}
\author{Stefan Scheel}
\email{s.scheel@imperial.ac.uk}
\affiliation{Quantum Optics and Laser Science, Blackett Laboratory,
Imperial College London, Prince Consort Road, London SW7 2BW, UK}
\author{Norbert L\"utkenhaus}
\affiliation{Institut f\"ur Theoretische Physik I,
Universit\"at Erlangen--N\"urnberg, Germany}
\affiliation{Max-Planck Research Group, Institute of Optics,
Information and Photonics, Universit\"at Erlangen--N\"urnberg, 
Staudtstr. 7/B3, D-91058 Erlangen, Germany}

\begin{abstract}
We develop an abstract way of defining linear-optics networks designed
to perform  quantum information tasks such as quantum gates. We
will be mainly concerned with the nonlinear sign shift gate, but it
will become obvious that all other gates can be treated in a similar
manner. The abstract scheme is extremely well suited for analytical as
well as numerical investigations since it reduces the number of
parameters for a general setting.
With that we show numerically and partially analytically for a wide
class of states that the success probability of generating a
nonlinear sign shift gate does not exceed $1/4$ which to our knowledge
is the strongest bound to date.
\end{abstract}

\date{\today}

\pacs{03.67.-a,42.50.-p,42.50.Ct}

\maketitle

%%%%%%%%%%%%%%%%%%%%%%%%%%%%%%%%%%%%%%%%%%%%%%%%%%%%%%%%%%%%%%%%%%%%%%
\section{Introduction}

One of the possible physical realisations of quantum information
processing is to use  conditional measurements in an all-optical passive
interferometric scheme. Within this framework, there are many
possibilities to encode qubits. Important examples are encodings in
superposition states of one photonic excitation in two modes
\cite{KLM}, in the polarization state of a photon
\cite{Koashi01,Pittman}, or in the occupation number states of one
mode.  In this paper, we will
follow the latter approach. Hence, we will denote by $|0\rangle$,
$|1\rangle$ etc. the vacuum state, the single-photon Fock state and so
on. A crucial part of building elementary quantum gates with this
encoding is the ability to perform a desired operation on two single
photons simultaneously. As an example, the controlled-$\sigma_z$
operation amounts to changing the sign of the two-mode state
$|11\rangle$ to $-|11\rangle$ thereby leaving the other three basis
states $|00\rangle$, $|01\rangle$ and $|10\rangle$ untouched. The unitary
operator associated with this quantum gate can be written in the form
$e^{i\pi\hat{n}_1\hat{n}_2}$ where the $\hat{n}_i$ are the number
operators of the photons in mode $i$. This unitary evolution operator
is quartic in the photonic amplitude operators and obviously stems
from a nonlinear interaction Hamiltonian with (scaled) `strength'
$\pi$. Apparently, neither fourth-order quantum electrodynamics nor
materials with effective Kerr nonlinearities reach the order of
magnitude required to perform the operation.

There is, however, an elegant way to circumvent the problem of
\textit{deterministically} generating the nonlinear evolution. The
idea is borrowed from the well-known theory of quantum-state
engineering (cf. Ref.~\cite{Welsch} and references cited
therein) where two optical fields in \textit{known} quantum states are
mixed at a beam splitter and postselection is performed subject to a
specific measurement outcome in the process of photo-detection in one
output arm. This idea has been generalised in \cite{KLM,Koashi01} to
\textit{unknown} quantum states in one input arm of the beam
splitter (which results in what one may call quantum-gate
engineering). The conditioning on a certain photo-detection event has
two consequences: (i) the \textit{effective} generation of a nonlinear
interaction Hamiltonian via postselection, and (ii) the
\textit{probabilistic} nature of the process due to the selection of a
subset of all possible measurement outcomes. We can equivalently call
this procedure the generation of measurement-induced nonlinearities
\cite{Scheel03}.

\begin{figure}[ht]
\centerline{\includegraphics[width=8cm]{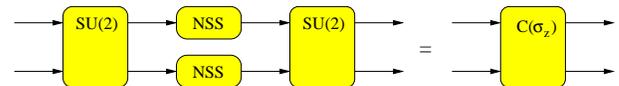}}
\caption{\label{fig:nss} Two nonlinear sign shift (NSS) gates inside a
balanced Mach--Zehnder interferometer are equivalent to a two-mode
controlled-$\sigma_z$ gate. Beam splitters are denoted by their group
action as SU($2$).}
\end{figure}

The particular example which has been elaborated in \cite{KLM} is the
nonlinear sign shift (NSS) gate which is described by the
transformation
\begin{equation}
\label{eq:nss}
c_0|0\rangle + c_1|1\rangle + c_2|2\rangle \mapsto
c_0|0\rangle + c_1|1\rangle - c_2|2\rangle \,.
\end{equation}
This NSS gate is intimately connected to the above-mentioned
controlled-$\sigma_z$ operation when noting that two NSS gates in each
arm of a balanced Mach--Zehnder interferometer is equivalent to a single
controlled-$\sigma_z$ gate (Fig.~\ref{fig:nss}).
Several different conditional measurement schemes have been found
that generate a NSS \cite{KLM,Scheel03,Ralph01}. In all cases, the
maximal success probability was found to be $1/4$. The question
that arises in this context is: what is the reason for this
particular value and is there a general upper limit on the
achievable success probability?
For the special case of a beam splitter network with
ancilla modes containing at most one photon each, it has been
shown analytically that the probability is bounded from above by
$1/2$ \cite{Knill03}. This bound is not necessarily tight since it
allows even for conditional dynamics, i.e. post-processing depending
on the outcome of a measurement. Note also that even then the success
probability does not reach unity.

In this article we introduce a compact model
for general linear optics networks. What we will  show with this
model is that in broad classes of conditional networks, the
success probability will not exceed the value of $1/4$ regardless
of the dimensionality of the ancilla state used. Although we do
not have a fully analytical proof for this claim, we will provide
strong numerical evidence and show steps towards a final proof.

This paper is organised as follows. In Sec.~\ref{sec:scheme} we
will describe the abstract conditional measurement scheme we will
investigate. This scheme is particularly simple but includes all
known networks so far. In section III the evaluation of the
success probability in the general case is prepared. The theory
of a general two-dimensional ancilla state is presented in
Sec.~\ref{sec:2d} followed by a discussion of special cases of
three-dimensional ancillas in Sec.~\ref{sec:3d}. In the last part,
we will give an outlook into possible future work in that
direction.

%%%%%%%%%%%%%%%%%%%%%%%%%%%%%%%%%%%%%%%%%%%%%%%%%%%%%%%%%%%%%%%%%%%%%%
\section{A general conditional measurement scheme}
\label{sec:scheme}

\begin{figure}[ht]
\centerline{\includegraphics[width=8cm]{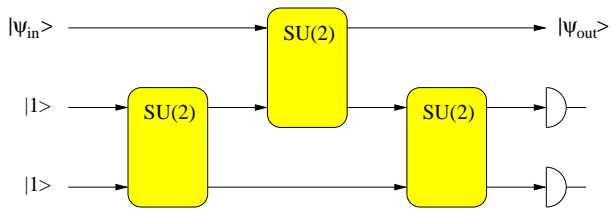}}
\caption{\label{fig:signshift} Beam splitter network realising the
nonlinear sign shift.}
\end{figure}

In this section we will describe an abstract conditional
measurement scheme that will allow us to investigate the success
probabilities of conditional optical networks without going into the detail of the specific
network. Throughout this article we will focus solely onto the NSS
gate, noting that other nonlinear gates can be treated analogously.
With single-photon sources and single-photon detectors it is known
that the NSS gate can be realised by the beam splitter network shown
in Fig.~\ref{fig:signshift} \cite{Scheel03}.
This network represents a group element of the unitary group
SU($3$) (note that a single lossless beam splitter acts as an element
of SU($2$) on the photonic amplitude operators
\cite{beamsplitter}).

\begin{figure}[ht]
\centerline{\includegraphics[width=8cm]{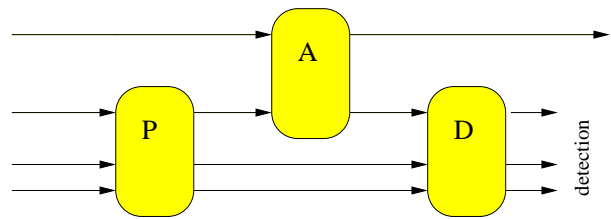}}
\caption{\label{fig:decomposition} Preparation stage `P' of the
ancilla state that is fed into the `active' beam splitter `A', and
decomposition stage `D' for detection.}
\end{figure}

It was noted some years ago that any U($N$)
group element can be realised by a triangle-shaped beam splitter
network \cite{Reck94}. Therefore, in order to generalise the network
in Fig.~\ref{fig:signshift} to incorporate an arbitrary number of
ancilla modes in arbitrary initial states, we
note that only a single beam splitter actually connects the
ancilla state(s) to the signal mode we want to act upon. This means
that there are several beam splitters that actually prepare a certain
ancilla state, feed them into this single `active' beam splitter which
is then followed by another set of beam splitters that prepare the
ancilla state for the detection process. Hence, we can schematically
represent the whole network by three blocks -- the preparation `P' of
the ancilla state from some given product states, the single `active'
beam splitter `A' connecting signal and ancilla modes, and the
detection stage `D' in which the ancilla, for some given initial
states, is again decomposed into product states (see
Fig.~\ref{fig:decomposition}).
This will be the general conditional measurement scheme we will be
looking at.

We can even go one step further by not requiring the preparation and
detection stages `P' and `D' to be realised by linear optical
elements. Instead, we will try to answer the question of upper bounds
for the success probability without this requirement. For that
purpose, we divide the total ancilla state into one mode that impinges
onto the `active' beam splitter `A' and mixes with the signal mode,
and the remaining modes.

From the general theory of measurement-induced nonlinearities we
know that, in order to realise an NSS gate, we need to generate an
operator which is a polynomial in the number operator
\cite{Scheel03}. The reasons for this constraint are on one hand
the requirement to produce an output state in a Hilbert space that
is isomorphic to the Hilbert space of the input state, and on the
other hand the necessity to be able to operate on each basis state
individually. We can rephrase these conditions, made from an
operator point of view, as a requirement on the ancilla state.
Isomorphism between the two Hilbert spaces of input and output
signal modes can only be achieved if the ancilla is restricted to
a state with fixed photon number and if the number of detected
photons equals the number of photons in the ancilla. In what
follows, we will therefore restrict our attention to ancilla
states with fixed photon numbers and the detection of the same number
of photons without losing generality.

From these observation it follows  that it suffices to consider
ancilla state containing exactly $N$ photons, which can hence be
written in the form
\begin{equation}
\sum\limits_{k=0}^N \gamma_k |k\rangle |A_{N-k}\rangle \,,
\end{equation}
where $|k\rangle$ is the $k$-photon Fock state impinging on `A'.
On the other hand, the state $|A_{N-k}\rangle$ can be any
multi-mode state as long as it contains exactly $N-k$ photons.
This means that we impose no restriction on the number of ancilla
modes. Neither have we said anything about the possibility of
creating this state with linear optical elements from suitable
product states, which is a question that can be addressed
separately. The $\gamma_k$ can be chosen to be real, since any
phase can be absorbed into the definition of $|A_{N-k}\rangle$,
and then fulfil the normalisation condition $\sum_k \gamma_k^2=1$.
The tensor-product state impinging onto `A' is thus
\begin{equation}
\Big( c_0|0\rangle+c_1|1\rangle+c_2|2\rangle \Big) \otimes
\sum\limits_{k=0}^N \gamma_k |k\rangle |A_{N-k}\rangle \,.
\end{equation}
So far, we have not said anything about the detection process itself,
we only select the contributions containing exactly $N$ photons. That
leaves us with a transformed state
\begin{eqnarray}
&& c_0 |0\rangle
\sum\limits_{k=0}^N \gamma_k |k\rangle |A_{N-k}\rangle (T^\ast)^k
\nonumber \\
&& +c_1|1\rangle
\sum\limits_{k=0}^N \gamma_k |k\rangle |A_{N-k}\rangle (T^\ast)^{k-1}
\left[ |T|^2-k|R|^2 \right] \nonumber \\
&& +c_2|2\rangle
\sum\limits_{k=0}^N \gamma_k |k\rangle |A_{N-k}\rangle (T^\ast)^{k-2}
\nonumber \\ && \times
\left[ |T|^4 -2k |T|^2 |R|^2+ \frac{k(k-1)}{2} |R|^4 \right] \nonumber \\
&=& c_0 \sqrt{N_0} |0\rangle|\psi_0\rangle
+ c_1 \sqrt{N_1} |1\rangle|\psi_1\rangle
+ c_2 \sqrt{N_2} |2\rangle|\psi_2\rangle \,. \nonumber \\
\end{eqnarray}
This state is yet unnormalised with its squared norm being equal
to the success probability.
Here, we have taken for `A' the conventional beam splitter matrix
\begin{equation}
\left( \begin{array}{cc} T & R \\ -R^\ast & T^\ast \end{array} \right)
\end{equation}
and have defined states $|\psi_l\rangle$ as
\begin{eqnarray}
\label{eq:psi}
|\psi_0\rangle &=& \frac{1}{\sqrt{N_0}}  \sum\limits_{k=0}^N
\gamma_k |k\rangle |A_{N-k}\rangle (T^\ast)^k  \,,\nonumber \\
|\psi_1\rangle &=& \frac{1}{\sqrt{N_1}} \sum\limits_{k=0}^N
\gamma_k |k\rangle |A_{N-k}\rangle (T^\ast)^{k-1}
\left[ |T|^2-k|R|^2 \right] \,,\nonumber \\
|\psi_2\rangle &=& \frac{1}{\sqrt{N_2}} \sum\limits_{k=0}^N \gamma_k
|k\rangle |A_{N-k}\rangle (T^\ast)^{k-2} \nonumber \\ && \times
\left[ |T|^4 -2k |T|^2 |R|^2+ \frac{k(k-1)}{2} |R|^4 \right]
\end{eqnarray}
and their corresponding normalisation factors
\begin{eqnarray}
N_0 &=& \sum\limits_{k=0}^N \gamma_k^2 |T|^{2k} \,, \nonumber \\
N_1 &=& \sum\limits_{k=0}^N \gamma_k^2 |T|^{2(k-1)}
\big[ (k+1)|T|^2-k \big]^2 \,,\nonumber \\
N_2 &=& \sum\limits_{k=0}^N \gamma_k^2 |T|^{2(k-2)}
\big[ |T|^4-2k|T|^2(1-|T|^2) \nonumber \\ &&
+\frac{k(k-1)}{2} (1-|T|^2)^2 \big]^2 \,.
\end{eqnarray}
As a matter of fact, we can rewrite the states $|\psi_l\rangle$ and
their respective norms in terms of Jacobi polynomials \cite{Gradstein}
noting that
\begin{eqnarray}
\lefteqn{
|nk\rangle \mapsto (T^\ast)^{k-n} {k \choose n}
\left( |T|^2-1 \right)^n } \nonumber \\ && \times
{}_2F_1\left[-n,-n,1+k-n;\frac{|T|^2}{|T|^2-1}\right] |nk\rangle
\nonumber \\ &=& (T^\ast)^{k-n} P_n^{(0,k-n)}(2|T|^2-1) |nk\rangle \,.
\end{eqnarray}

The actual measurement consists of a
single projection onto some state $\langle\psi|$ which, with a certain
probability, gives us the desired signal state. The conditions we have
to fulfil are
\begin{equation}
\label{eq:conditions}
\sqrt{N_0}\langle\psi|\psi_0\rangle =
\sqrt{N_1}\langle\psi|\psi_1\rangle =
-\sqrt{N_2}\langle\psi|\psi_2\rangle \,.
\end{equation}
Note that the states $|\psi_l\rangle$ depend (in a rather complicated
way) on the complex transmission coefficient $T$ and the weights
$\gamma_k$ only.
Assuming that the condition (\ref{eq:conditions}) is fulfilled,  the probability that
the measurement outcome corresponds to the projection onto
$\langle\psi|$ can be written as
\begin{equation}
p_{\text{success}} =  N_0 \left|\langle\psi|\psi_0\rangle\right|^2 \,.
\end{equation}
Remember that the state $\langle\psi|$  has to be determined by the
conditions (\ref{eq:conditions}). In the following we will explore the implications of these conditions
in settings that depend on the dimension of the state space spanned by
the ancilla.

%%%%%%%%%%%%%%%%%%%%%%%%%%%%%%%%%%%%%%%%%%%%%%%%%%%%%%%%%%%%
\section{Fully three-dimensional ancilla space}

In the three-dimensional space spanned by the $|\psi_l\rangle$, we can
choose a basis in which $|\psi_0\rangle$ is represented by the vector
$(1,0,0)^T$. With the help of the Gram--Schmidt orthogonalisation
procedure we define the representation vectors for $|\psi_1\rangle$
and $|\psi_2\rangle$ as
\begin{equation}
\left( \begin{array}{c}   y_1 \\ y_2 \\ 0 \end{array} \right)
\quad \mbox{and} \quad
\left( \begin{array}{c}   z_1 \\ z_2 \\ z_3 \end{array} \right) \,,
\end{equation}
respectively. Here we have the interpretations $y_1 =
\langle\psi_0|\psi_1\rangle$, $z_1 = \langle\psi_0|\psi_2\rangle$ and
$z_2 = \frac{\langle\psi_1|\psi_2\rangle-y_1^\ast z_1}{y_2}$
while $y_2$ and $y_3$ can be chosen to be real numbers according to
the required normalization.
In this section we discuss the scenario where the
ancilla space is indeed three-dimensional, so that $y_2 \neq 0$ and $
z_3 \neq 0$. Then,  we can define a representation for the detected state
$\langle\psi|$ as $(\alpha,\beta,\gamma)^T$ such that the success
probability can be given in the compact form
\begin{equation}
p_{\text{success}} = |\alpha|^2 N_0 \,.
\end{equation}
Remember that the state $\langle\psi|$ and therefore the value of
$\alpha$ has to be determined by the conditions (\ref{eq:conditions}).
Inserting the representations for the $|\psi_l\rangle$ into
(\ref{eq:conditions}) leaves us successively with
\begin{eqnarray}
\beta & = & \alpha
\frac{1}{y_2}\left(\sqrt{\frac{N_0}{N_1}}-y_1^\ast\right) \\
\gamma&=& -\frac{\alpha}{z_3} \left[ \sqrt{\frac{N_0}{N_2}}+z_1^\ast
+\frac{z_2^\ast}{y_2}\left(\sqrt{\frac{N_0}{N_1}}-y_1^\ast\right)\right].
\end{eqnarray}
The required normalisation condition
$1=|\alpha|^2+|\beta|^2+|\gamma|^2$ then leads to
an expression for the success probability
in terms of the coefficients of the representation vectors and the
normalisations only,
\begin{eqnarray}
\label{eq:fullprob}
\lefteqn{
p_{\text{success}} = N_0 \Bigg[ 1
+\frac{1}{|y_2|^2} \left| \sqrt{\frac{N_0}{N_1}}-y_1 \right|^2 }
\nonumber \\ &&
+\frac{1}{|z_3|^2} \left| \sqrt{\frac{N_0}{N_2}}+z_1
+\frac{z_2}{y_2}\left(\sqrt{\frac{N_0}{N_1}}-y_1\right) \right|^2
\Bigg]^{-1} \,.
\end{eqnarray}
There may be many different solutions with different success
probabilities. Our task is now to find suitable upper bounds.
With the general theory presented above, we are now able to look into
certain special cases where the Hilbert space dimension of the ancilla
state is sufficiently low.

%%%%%%%%%%%%%%%%%%%%%%%%%%%%%%%%%%%%%%%%%%%%%%%%%%%%%%%%%%%%%%%%%%%%%%
\section{Two-dimensional ancillas}
\label{sec:2d}

An interesting special case occurs when the ancilla is only
two-dimensional. This can happen, for example, if the total photon
number is only  one or if only two of the $\gamma_k$ are non-zero.
For example, the nonlinear sign shift gate in \cite{Ralph01} employs
an initial ancilla state of the form $|01\rangle$ which, after feeding
it into the preparation step `P' of a network similar to
Fig.~\ref{fig:nss} (which is just a single beam splitter), produces a
two-dimensional ancilla $\alpha|01\rangle+\beta|10\rangle$, one mode
of which mixes with the signal mode at the beam splitter `A'.

In such a case, the states $|\psi_l\rangle$ are not independent of
each other. In the two-dimensional space spanned by the
$|\psi_l\rangle$, we can then choose a basis in which $|\psi_0\rangle$
is represented by the vector $(1,0)^T$. With the help of the
Gram--Schmidt orthogonalisation procedure we define the representation
vectors for $|\psi_1\rangle$ and $|\psi_2\rangle$ as
\begin{equation}
\left( \begin{array}{c}   y_1 \\ y_2  \end{array} \right)
\quad \mbox{and} \quad
\left( \begin{array}{c}   z_1 \\ z_2  \end{array} \right) \,,
\end{equation}
respectively, where
\begin{eqnarray}
y_1 &=& \langle\psi_0|\psi_1\rangle \,,\nonumber \\
y_2 &=& \sqrt{1-|y_1|^2} \,,\nonumber \\
z_1 &=& \langle\psi_0|\psi_2\rangle \,, \nonumber \\
z_2 &=& \frac{\langle\psi_1|\psi_2\rangle-y_1^\ast z_1}{y_2}\; .
\end{eqnarray}
With this notation, we find that the conditions (\ref{eq:conditions})
lead to two different constraints: one condition
leads to a relationship between $\alpha$ and $\beta$, and therefore  via
the normalization condition again to an expression for the success
probability,
\begin{equation}
\label{eq:prob2}
p_{\text{success}} = N_0 \left( 1+ \frac{1}{|y_2|^2} \left|
\sqrt{\frac{N_0}{N_1}}-y_1 \right|^2 \right)^{-1} \,.
\end{equation}
 The second condition, that related in the
three-dimensional case $\beta$ and $\gamma$, now takes the form
\begin{equation}
\label{eq:cond2}
-\sqrt{\frac{N_0}{N_2}} = z_1 + \frac{z_2}{y_2} \left(
\sqrt{\frac{N_0}{N_1}} - y_1 \right)\; .
\end{equation}

In order to evaluate the achievable success probability under these
constraints, let us now specify the ancilla state. As we have already
noted, this can be achieved if only
two weight factors are non-zero. Thus, the ancilla has the form
\begin{equation}
\gamma |m\rangle|A_n\rangle + \sqrt{1-\gamma^2}|n\rangle|A_m\rangle \,.
\end{equation}
Hence, we assume that $N=m+n$ but not all possible superpositions are
allowed for $N>1$. Remember also that $\gamma$ can be chosen to be real.
As a consequence, the constraint (\ref{eq:cond2}) can be written in
terms of Jacobi polynomials as
\begin{eqnarray}
&& T^\ast
\left[ P_0^{(0,m)} P_2^{(0,n-2)} -P_2^{(0,m-2)} P_0^{(0,n)} \right]
\nonumber \\ && +(T^\ast)^2
\left[ P_0^{(0,m)} P_1^{(0,n-1)} -P_1^{(0,m-1)} P_0^{(0,n)} \right]
\nonumber \\ & = &
\left[ P_1^{(0,m-1)} P_2^{(0,n-2)} -P_2^{(0,m-2)} P_1^{(0,n-1)} \right]
\,,
\end{eqnarray}
where we omitted the common arguments $(2|T|^2-1)$ in all Jacobi
polynomials.  This equation is  independent of the weight $\gamma$ and
takes the form of a quartic equation
\begin{eqnarray}
\label{eq:quartic}
\lefteqn{
|T|^4 (m+1)(n+1) -|T|^2T^\ast(m+n+3) -2T^{\ast 2} }
\nonumber \\ && \hspace*{-3ex}
-|T|^2\left( m+n+2mn -1\right)
+T^\ast(m+n-1) +mn =0 \,. \nonumber \\
\end{eqnarray}
The success probability (\ref{eq:prob2}),
however, then depends strongly on $\gamma$. Performing the calculation
leads to an expression for the success probability as
\begin{equation}
\label{eq:prob2a}
p_{\text{success}} = \frac{(m-n)^2(1-|T|^2)^2}
{\frac{|T^\ast-(n+1)|T|^2+n|^2}{\gamma^2 T^{2m}}
+\frac{|T^\ast-(m+1)|T|^2+m|^2}{(1-\gamma^2)T^{2n}}} \,.
\end{equation}
From this expression we already see that neither $T=0$ nor $T=\pm1$ give
a non-zero probability.
Another  fact to notice is that one-dimensional ancillas ($m=n$) give
zero probability, too, which was to be expected. The maximum of
Eq.~(\ref{eq:prob2a}) is reached for
\begin{equation}
\gamma_{\text{max}}^2 = \left[ 1 \pm \left|
\frac{|T|^m[T^\ast-(m+1)|T|^2+m]}{|T|^n[T^\ast-(n+1)|T|^2+n]} \right|
\right]^{-1}
\end{equation}
where the sign has to be chosen such that
$0\le\gamma_{\text{max}}^2\le 1$. The corresponding success
probability reads
\begin{widetext}
\begin{equation}
\label{eq:maxprob2}
p^{\text{max}}_{\text{success}} =
\frac{ (m-n)^2 (1-|T|^2)^2 |T|^{2(m+n)}}
{\left[ |T|^m[T^\ast-(m+1)|T|^2+m]+ |T|^n [T^\ast-(n+1)|T|^2+n]\right]^2}
\end{equation}
\end{widetext}
where the solution from Eq.~(\ref{eq:quartic}) has to be inserted for $T$.

%%%%%%%%%%%%%%%%%%%%%%%%%%%%%%%%%%%%%%%%%%%%%%%%%%%%%%%%%%%%%%%%%%%%%%
\subsection{Special cases}

There are certain special cases which we can investigate
analytically. First of all, let us take the transmission
coefficient to be real, $T\in\mathbb{R}$. Then, the quartic
equation (\ref{eq:quartic}) reduces to
\begin{eqnarray}
\label{eq:quartic2}
\lefteqn{
T^4 (m+1)(n+1)-T^3(m+n+3) } \nonumber \\ &&
-T^2( m+n+2mn +1) +T(m+n-1) +mn =0 \,. \nonumber \\
\end{eqnarray}
Furthermore, let us restrict ourselves to the case when the photon
numbers are adjacent integers, hence $m=n+1$. Then, we find solutions
to Eq.~(\ref{eq:quartic2}) as
\begin{equation}
T_{1,2} = \pm \sqrt{\frac{n}{n+2}} \,,\quad T_3 =
\frac{1-\sqrt{n^2+2n+2}}{n+1} \,.
\end{equation}
The fourth solution is greater than $1$ for all values of $n$ and has
to be omitted. Inserting these solutions into Eq.~(\ref{eq:maxprob2}),
we obtain
\begin{equation}
p^{\text{max}}_{\text{success}}(T_{1,2}) = \frac{1}{4} \left(
\frac{n}{n+2} \right)^n \left( \sqrt{\frac{n}{n+2}} \pm 1 \right)^2
\end{equation}
as well as
\begin{equation}
p^{\text{max}}_{\text{success}}(T_3) = \frac{z}{(\sqrt{x}+\sqrt{y})^2}
\end{equation}
with
\begin{eqnarray}
x &=& 3+n+2n^2+n^3+n^4 \nonumber \\ && -(n^3+n-2)\sqrt{n^2+2n+2} \,,
\nonumber \\
y &=& (1+n)^2(3+3n+n^2-(n+2)\sqrt{n^2+2n+2}) \,, \nonumber \\
z &=& 2(n+1)^{2(1-n)} (1-\sqrt{n^2+2n+2})^{2n} \,.
\end{eqnarray}
As functions of $n$, the functions $x(n)$, $y(n)$ and $z(n)$ are all
monotonically increasing, with $x(n)$ and $y(n)$ behaving as
$\propto n^3$ and $z(n)\propto n^2$. Hence,
$p^{\text{max}}_{\text{success}}(T_3)$ is monotonically decreasing
with $n$ and therefore attains its maximum value for $n=0$. For all
three solutions we then obtain the final answer for this special class
as
\begin{equation}
p^{\text{max}}_{\text{success}} \le \frac{1}{4} \,.
\end{equation}

A very special, and well-known, case is realised when $N=1$. For
example, if the initial ancilla product state is chosen as
$|1\rangle|0\rangle$ from which a general ancilla state can be
generated by a beam splitter, the total photon number is just
$N=1$. Then, we can set $m=0$ and $n=1$, and Eq.~(\ref{eq:quartic})
reduces to
\begin{equation}
T^2 (T^2-2T-1) = 0
\end{equation}
with the solutions $T=1-\sqrt{2}$ and $T=0$.
The first solution is precisely the
one known from the literature \cite{Scheel03,KLM,Ralph01} and leads to
a maximal success probability of 1/4 which is known to be optimal for
this situation. The second solution has vanishing
probability.

Furthermore, as one can see from Eq.~(\ref{eq:quartic}),
whenever we have $m=0$, one of the solutions is $T=0$. However, all
these solutions have vanishing success probability. From the remaining
cubic equation,
\begin{equation}
T^3-\frac{n+3}{n+1}T^2-T+\frac{n-1}{n+1}=0 \,,
\end{equation}
we see that there are simple solutions in the (open) intervals
$(-1,0)$ and $(0,1)$. However, the expressions for their corresponding
success probabilities are too complicated to warrant reproduction in
this article. Instead, we will resort to numerical computation.

\begin{figure}[ht]
\centerline{\includegraphics[width=7.5cm]{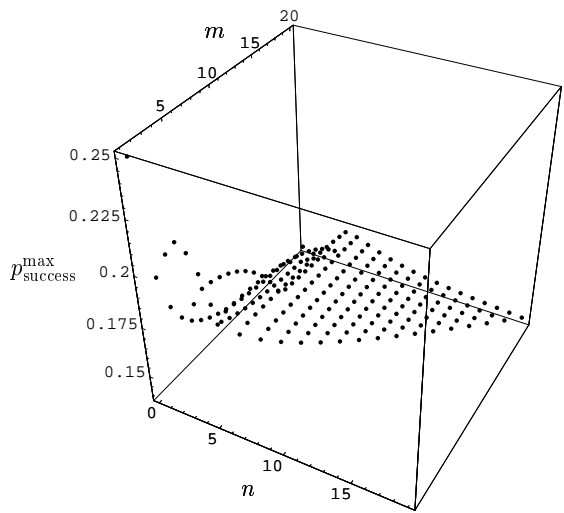}}
\caption{\label{fig:prob2} Maximal success probability for different
values of $m$ and $n$ upto 20. The probability never exceeds 1/4. In
this figure, we have used the symmetry with respect to the interchange
$m\leftrightarrow n$ to discard half the results.}
\end{figure}

In Fig.~\ref{fig:prob2} we show the success probabilities for
different pairs of integers ($m,n$) where we assume that the
transmission coefficient $T$ can be taken to be real. The probability
never exceeds $1/4$, and this value is only reached for $m=0$, $n=1$
as shown above. 
Note that this result is independent of the number of ancilla
modes used, as long as they represent only a two-dimensional
space. That is, ancillas that are produced from product states of
the form $|1\rangle|0\rangle\cdots|0\rangle$ do not give an
improvement either. Note also that we still did not require the
ancilla state to be producable from product states by linear
optics, though in the example considered above this can be done
that way. That is, the mere requirement to measure exactly $N$
photons after the interaction between signal and ancilla with a
beamsplitter seems to give a bound on the success probability.

%%%%%%%%%%%%%%%%%%%%%%%%%%%%%%%%%%%%%%%%%%%%%%%%%%%%%%%%%%%%%%%%%%%%%%
\section{Example of a three-dimensional ancilla state}
\label{sec:3d} 

A slightly more general situation is
encountered if we allow for three-dimensional ancilla states. An
example for this case is provided by the beam splitter network
depicted in Fig.~\ref{fig:signshift} with single-photon ancillas.
Let us denote the unitary (3$\times$3)-matrix associated with the
network by $\bm{\Lambda}$, hence its matrix elements are given by
$\Lambda_{ij}$. Then it has been shown in Ref.~\cite{Scheel03}
that the success probability is given by
\begin{equation}
\label{eq:permanent} p_{\text{success}} =
\frac{4|\Lambda_{12}\Lambda_{13}\Lambda_{21}\Lambda_{31}|^2}
{|1+2\Lambda_{11}-\Lambda_{11}^2|^2} \,.
\end{equation}
 Numerically, a maximal value of
$p_{\text{success}}\approx 0.235$ has been found in
\cite{Scheel03}.

Let us now see what bounds we can derive from our theory. As a matter
of fact, we can already find an upper bound from unitarity of
$\bm{\Lambda}$. Because the network is unitary, we have the inequality
\begin{equation}
\label{eq:lemma}
|\Lambda_{12}\Lambda_{13}\Lambda_{21}\Lambda_{31}| \le \frac{1}{4}
\left( 1-|\Lambda_{11}|^2 \right)^2 \,.
\end{equation}
To see this, note that
$|\Lambda_{11}|^2+|\Lambda_{12}|^2+|\Lambda_{13}|^2=1$
which permits the parametrisation $|\Lambda_{11}|=|\cos\alpha|$,
$|\Lambda_{12}|=|\sin\alpha\cos\beta|$, and
$|\Lambda_{13}|=|\sin\alpha\sin\beta|$. Therefore,
\begin{eqnarray}
|\Lambda_{12}\Lambda_{13}| &=& |\sin^2\alpha \sin\beta\cos\beta|
\nonumber \\ &=& \frac{1}{2} |\sin 2\beta| (1-|\Lambda_{11}|^2) \,.
\end{eqnarray}
Since $|\sin\beta|\le 1$, we immediately find that
$|\Lambda_{12}\Lambda_{13}|\le (1-|\Lambda_{11}|^2)/2$.
An analogous relation holds for the product $|\Lambda_{21}\Lambda_{31}|$
and relation (\ref{eq:lemma}) is proven.

Thus, the success probability is bounded from above by
\begin{equation}
\label{eq:bound3}
p_{\text{success}} \le
\frac{(1-|\Lambda_{11}|^2)^4}{4|1+2\Lambda_{11}-\Lambda_{11}^2|^2}
\end{equation}
which only depends on one (complex) parameter,
$\Lambda_{11}$. Since the solution for $\Lambda_{11}$ can be taken
to be phase-independent if we allow for a global phase in the final signal
state, we have to minimize the bound with respect to the phase for
fixed magnitude of $\Lambda_{11}$. Then it turns out that
$\Lambda_{11}=0$ is just the sought minimum leaving us with an upper
bound of 1/4 just as in the case of a two-dimensional ancilla.

Let us now look what implications the theory has which we have
developed in the previous sections. We consider a
three-dimensional ancilla state of the form
\begin{equation}
\gamma_1 |0\rangle|A_N\rangle +\gamma_2 |1\rangle|A_{N-1}\rangle
+\sqrt{1-\gamma_1^2-\gamma_2^2} |2\rangle|A_{N-2}\rangle \;.
\end{equation}
where we take $\gamma_1,\gamma_2\in\mathbb{R}$. Hence, we assume that
only states with at most two photons mix with the signal state
which is certainly only a special case.
A particular example would be just the situation considered above
where we started with a product state of the form $|1\rangle|1\rangle$
which has a total photon number of two. However, the abstract setting
developed in this paper allows to treat all three-dimensional ancillas
simultaneously without referring to a specific network.
This demonstrates that our analysis allows to search readily larger
classes of strategies for good strategies.

\begin{figure}[th]
\centerline{\includegraphics[width=8cm]{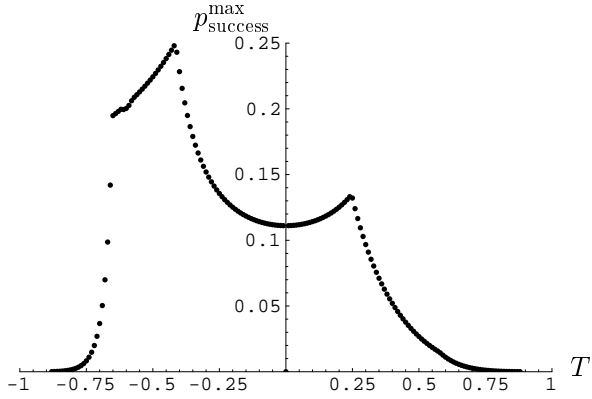}}
\caption{\label{fig:prob3} Maximal success probability for a
three-dimensional ancilla state with at most two photons impinging
onto the active beam splitter for real $T$.}
\end{figure}

Suppose now in addition that the transmission coefficient $T$ is
real. If we  vary the magnitude of the transmission coefficient,
we obtain the results shown in Fig.~\ref{fig:prob3}.
What we can see is that the success probability reaches $1/4$
at one point. Incidentally, there the transmission coefficient obtains
the value $T=1-\sqrt{2}$ which is precisely the value found for
two-dimensional ancillas with a total photon number of one. In fact,
this is not too surprising. A closer look at Fig.~\ref{fig:prob3}
reveals that there are three local maxima. Each of them is located at
the value of $T$ that is a solution of the two-dimensional ancilla
problem with one of the photon numbers absent due to destructive
interference. The reason for the appearance of these maxima can be
understood by going back to the original expression,
Eq.~(\ref{eq:fullprob}). Note that the denominator in there consists
of three non-negative numbers. One of those terms being zero certainly
increases the success probability. A closer look reveals that for the
last term,
\begin{equation}
\sqrt{\frac{N_0}{N_2}}+z_1
+\frac{z_2}{y_2}\left(\sqrt{\frac{N_0}{N_1}}-y_1\right) \,,
\end{equation}
to be zero, condition (\ref{eq:cond2}) has to be satisfied. But this
is just the condition that gives the solution for $T$ in case of a
two-dimensional ancilla state.

Now we have an interesting
situation at hand. We found that the maximal success probability
for a class of three-dimensional ancilla states does not increase
beyond the maximal value obtained by using just a two-dimensional
ancilla. Moreover, the minimum ist achieved for a parameter choice
where one of the three initial photon number contributions
disappears due to destructive interference. From this observation
we conjecture that adding a third dimension to the ancilla Hilbert
space does not lead to any improvement in success probability.
This adds to the fact that the number of ancilla modes does not
matter either.

%%%%%%%%%%%%%%%%%%%%%%%%%%%%%%%%%%%%%%%%%%%%%%%%%%%%%%%%%%%%%%%%%%%%%%
\subsection{A geometric view of the success probability}

Here we introduce yet another way of viewing the probability. For that,
let us define unnormalised states
$|\phi_0\rangle=\sqrt{N_0}|\psi_0\rangle$,
$|\phi_1\rangle=\sqrt{N_1}|\psi_1\rangle$, and
$|\phi_2\rangle=\sqrt{N_2}|\psi_2\rangle$ with the following overlaps
and norms:
\begin{eqnarray}
v_{01} &=& \langle\phi_0|\phi_1\rangle \,,\quad
N_0 = \langle\phi_0|\phi_0\rangle \,,\nonumber \\
v_{02} &=& \langle\phi_0|\phi_2\rangle \,,\quad
N_1 = \langle\phi_1|\phi_1\rangle \,,\nonumber \\
v_{12} &=& \langle\phi_1|\phi_2\rangle \,,\quad
N_2 = \langle\phi_2|\phi_2\rangle \,.
\end{eqnarray}
With this notation, the success probability (\ref{eq:fullprob}) can be
written in the following form (at least for $T\in\mathbb{R}$):
\begin{widetext}
\begin{eqnarray}
p_{\text{success}} &=&
\frac{(N_0v_{12}^2+N_1v_{02}^2+N_2v_{01}^2)-N_0N_1N_2-2v_{01}v_{02}v_{12}}%
{2(v_{01}v_{02}+v_{01}v_{12}-v_{02}v_{12})-2(N_0v_{12}+N_1v_{02}-N_2v_{01})
+(v_{01}^2+v_{02}^2+v_{12}^2)-(N_0N_1+N_0N_2+N_1N_2)}
\nonumber \\ &=&
\frac{(N_0v_{12}^2+N_1v_{02}^2+N_2v_{01}^2)-N_0N_1N_2-2v_{01}v_{02}v_{12}}%
{(v_{01}+v_{02}+v_{12})^2-4v_{02}v_{12}-2(N_0v_{12}+N_1v_{02}-N_2v_{01})
-(N_0N_1+N_0N_2+N_1N_2)} \,.
\end{eqnarray}
\end{widetext}
Interestingly, the only changes compared with the situation in which
we intend to perform the identity operation is that we make the
replacements $v_{02}\mapsto-v_{02}$, $v_{12}\mapsto-v_{12}$. In this
case the maximal probability reaches unity for $T=1$.

If we go back to the vector notation for the representation vectors
for the unnormalised states $|\phi_0\rangle$, $|\phi_1\rangle$,
and $|\phi_2\rangle$ and assume them to be real, we find that we can
rewrite the success probability in the form
\begin{equation}
p_{\text{success}} =
\frac{\left[(\phi_0\times\phi_1)\cdot\phi_2\right]^2}%
{\left[(\phi_0\times\phi_1)-\phi_2\times(\phi_0-\phi_1)\right]^2}
\end{equation}
where the minus sign between the two terms in the denominator is the
one from the nonlinear sign shift. This kind of representation allows
for geometric interpretation. For example, let us introduce the dual
vectors
\begin{eqnarray}
\phi_0' &=& \phi_1 \times \phi_2 \,,\\
\phi_1' &=& \phi_2 \times \phi_0 \,,\\
\phi_2' &=& \phi_0 \times \phi_1 \,.
\end{eqnarray}
Then the success probability reduces to
\begin{equation}
p_{\text{success}} =
\frac{(\phi_0'\times\phi_1')\cdot\phi_2'}{|\phi_0'+\phi_1'-\phi_2'|^2}
\end{equation}
which is nothing but the ratio between the volume of the
parallelepiped spanned by $\phi_0'$, $\phi_1'$, and $\phi_2'$, and its
squared diagonal length. Note that in case we would do nothing we
would have to convert the `$-$' sign in the denominator into a `$+$'.

We have not found a suitable way to exploit this relationship, but we
believe that it can be a key for more general results.

%%%%%%%%%%%%%%%%%%%%%%%%%%%%%%%%%%%%%%%%%%%%%%%%%%%%%%%%%%%%%%%%%%%%%%
\section{Conclusion and discussion}

In this article we have presented an abstract way of looking at
linear-optics quantum information networks. Thereby we reduce the
network to a single `active' beam splitter and disregard all
details about the actual ancilla-state preparation and detection
stages. We believe this to be a step towards a fully analytical
proof of upper bounds for success probabilities for which we have
only numerically support so far. In our analysis we have focussed
onto an important type of gate, namely the nonlinear sign shift
which is the archetypal probabilistic single-mode gate. From
previous research several restrictions on the structure of the
ancilla state and the detection process could be made. We have
shown semi-analytically and numerically that for wide classes of
ancilla states the highest achievable success probability is
$1/4$. This leads us to the conjecture that this value represents
a bound valid for all implementations of the nolinear sign shift
gate.  This conjectured bound is tight in a way that it can
actually be realised. Several examples for different networks
exist in the literature.

From the structure of the problem it seems that for this type of
low-dimensional states we encountered here, a fully analytical proof
for the upper bound should be possible in the near
future. Additionally, the structure of problem will also allow to
study the scaling behaviour of the probability with increasing
dimension of the signal mode, that is, if we were to look at
generalized nonlinear sign shift gates. That, however, will be the
subject of future investigations.

A further remark concerns the application of our result to higher-mode
signal states. As mentioned earlier, in order to realize a
controlled-$\sigma_z$ operation, in principle one could place two
nonlinear sign shift gates inside a balanced Mach--Zehnder
interferometer (Fig.~\ref{fig:nss}). Since each of the NSS gate works
only with a success probability of $1/4$, the controlled-$\sigma_z$
gate would then work with a probability of $1/16$. However, a scheme
is already known that works with a probability of $2/27$
\cite{Knill02}. This result points towards a nontrivial multimode
extension of our scheme in which not only one beam splitter is
`active'.

In the present article, we have focussed onto the dimensionality
of the ancilla state that interacts with the single-mode signal
state at the `active' beam splitter. We have found that the
often-quoted intuition that an increase of the auxiliary
dimensions improves the performance of a linear-optics gate
implementation may not hold. In fact, supported by further
preliminary studies, we conjecture that an ancilla with its
dimension being reduced by one with respect to the dimension of
the  signal state  is sufficient.  Not only is the number of
ancilla modes irrelevant to our scheme, also the Hilbert space
dimension of the ancilla state impinging onto the `active' beam
splitter can be kept to a minimum. Indeed, the bound of $1/4$ is
already achieved with a two-dimensional ancilla containing only at
most one photon.

This astonishing result has important consequences
for the practicability and realisability of such networks. From a
theoretical point of view, low photon numbers in a network means low
decoherence. It is well-known that Fock states with higher photon
numbers (being more `nonclassical' with respect to some measure that
needs to be defined) are much more fragile than their low-number
counterparts. On the other hand, the low dimension of the ancilla and
low photon numbers will keep experimental resources at a minimum. We
believe that our result is encouraging to warrant more research in the
area of linear-optics quantum information processing.

\acknowledgments This project was partly funded by the Deutsche
Forschungsgemeinschaft (DFG) under the Emmy-Noether Programme, the
UK Engineering and Physical Sciences Research Council (EPSRC), and
the European Commission (RAMBOQ programme). One of us (S.S.)
kindly acknowledges support by the Alexander~von~Humboldt
foundation through the Feodor-Lynen program and support by the
European Science Foundation.

%%%%%%%%%%%%%%%%%%%%%%%%%%%%%%%%%%%%%%%%%%%%%%%%%%%%%%%%%%%%%%%%%%%%%%

\end{document}